%% file: conference_paper.tex
\documentclass[conference,a4paper,10pt]{IEEEtran}
\IEEEoverridecommandlockouts

\usepackage[T1]{fontenc} 
\usepackage[utf8]{inputenc}
\usepackage[english]{babel}
\usepackage{glossaries}
\input{acronym}
\usepackage{lmodern}
\usepackage{comment}
\usepackage[colorlinks=true, allcolors=blue]{hyperref}
\usepackage{extarrows}
\usepackage{cite}
\usepackage[normalem]{ulem}
\usepackage[svgnames]{xcolor}
\usepackage[pdftex]{graphicx}
\usepackage{caption}
\usepackage{subcaption}
\usepackage{tikz}
\usetikzlibrary{math, shapes}
\usepackage{pgfplots}
\pgfplotsset{
compat=1.3,
legend style={font=\scriptsize, fill opacity=0.8,  draw opacity=1, text opacity=1, draw=white!15!black, legend cell align=left, align=left}, 
width=6.5cm, 
yminorticks=false,
xminorticks=false,
title style={font=\small},
tick style={color=black},
tick label style={font=\small},
tick align=outside,
grid style={line width=.1pt, draw=gray!20},
major grid style={line width=.1pt,draw=gray!20},
label style={font=\small},
}
\usepackage{amssymb,amsmath,amsfonts,amsthm,bm,mathtools,cuted,bbold}
\usepackage[cmintegrals]{newtxmath}
\usepackage{cuted}
\usepackage{tabularx}
\usepackage{booktabs}
\usepackage[ruled,vlined,linesnumbered]{algorithm2e}

\newcommand{\T}{^{\intercal}}     
\newcommand{\E}[1]{\mathbb{E}\left\{ #1 \right\}} 
\newcommand{\mc}[1]{\mathcal{#1}}   
\newcommand{\mb}[1]{\mathbf{#1}}    
\DeclareMathOperator*{\argmax}{arg\,max}    
\DeclareMathOperator*{\argmin}{arg\,min}    
\DeclareMathOperator{\sinc}{sinc}

\newcommand{\HPp}{^{\rm \scriptscriptstyle {HP+}}}
\newcommand{\HPm}{^{\rm \scriptscriptstyle {HP-}}}

\definecolor{gold}{rgb}{0.85,.66,0}
\definecolor{amaranth}{rgb}{0.9, 0.17, 0.31}

\newtheorem{assumption}{Assumption}
\newtheorem{lemma}{Lemma}

\title{Localization-based OFDM framework for RIS-aided systems}
\author{\IEEEauthorblockN{Fabio Saggese\IEEEauthorrefmark{1}, Kimmo Kansanen\IEEEauthorrefmark{1}\IEEEauthorrefmark{2}, and Petar Popovski\IEEEauthorrefmark{1}}
\IEEEauthorblockA{\IEEEauthorrefmark{1}Department of Electronic Systems, Aalborg University, Denmark (\{fasa, kimkan, petarp\}@es.aau.dk)}
\IEEEauthorblockA{\IEEEauthorrefmark{2}Department of Electronic Systems, Norwegian University of Science and Technology}\thanks{This work was partly supported by the Villum Investigator grant ``WATER'' from the Villum Foundation, Denmark, and by the Horizon 2020 ``RISE-6G'' project, financed by the European Commission under grant no. 101017011.}}

\begin{document}

\maketitle

\begin{abstract}
Efficient integration of reconfigurable intelligent surfaces (RISs) into the current wireless network standard is not a trivial task due to the overhead generated by performing channel estimation (CE) and phase-shift optimization. In this paper, we propose a framework enabling the coexistence between orthogonal-frequency division multiplexing (OFDM)
and RIS technologies. Instead of wasting communication symbols for the CE and optimization, the proposed framework exploits the localization information obtainable by RIS-aided communications to provide a robust
allocation strategy for user multiplexing. The results demonstrate the effectiveness of the proposed approach with respect to CE-based transmission methods.
\end{abstract}

\section{Introduction}
\Glspl{RIS} are considered among the enabling technologies for the next-generation (6G) of wireless networks due to their ability to control the wireless propagation environment~\cite{bjornson2021signalprocessing}. 

In recent years, many research works have shown the capabilities of the \gls{RIS} in terms of improved coverage, data rate, and mitigation of multi-user interference through accurate phase-shift optimization~\cite{mengnan2022survey}. 
However, only a few works have focused on integrating the \gls{RIS} into the existing communication protocols. To the best of the authors' knowledge, the authors of~\cite{Zheng2020cebf} were the first to consider \gls{RIS}-aided communications framework employing \gls{OFDM}, proposing a joint \gls{CE} and \gls{RIS} phase-shift optimization framework showing outstanding performance for a single \gls{UE} scenario.
Nevertheless, in multi-user systems, \gls{RIS} channel estimation procedures need pilot sequences whose length is proportional to the number of \gls{RIS} elements and the number of \glspl{UE} in the system~\cite{Wang2020ce, mengnan2022survey}. Hence, the \gls{CE} procedure may lead to an unfeasible overhead when serving a large number of users.

From a different perspective, the recent literature has shown the capability of using \gls{RIS} to enhance the performance of localization algorithms~\cite{bjornson2021signalprocessing}, and even enable new localization procedures not feasible without an \gls{RIS}~\cite{keykhosravi2021siso}.
Considering the relevance that \gls{ISAC} has gained to enable the \emph{connected intelligence} promised by the 6G, it is natural to imagine the use of the (improved) localization information available in \gls{RIS}-aided networks to help the scheduling decision: instead of wasting time for exchanging \gls{CE} pilot sequences, the knowledge of the \glspl{UE} position can be used to optimize the data transmission.
In this line, the authors of~\cite{Abrardo2021positioning} proposed a \gls{RIS} phase shift optimization based on localization information able to maximize the average spectral efficiency. Furthermore, in~\cite{wang2021joint}, the authors provide a \gls{TDD} framework to integrate communication and localization in \gls{RIS}-aided systems.

This paper proposes an \gls{OFDM} scheduling protocol for \gls{RIS}-aided systems based on localization information. 
 Following the time horizon provided by the \gls{OFDM} structure, our \gls{RIS} loads different configurations in different time slots, enhancing incoming signals in a space division manner. The configurations are stored in a \emph{codebook}, optimized to provide maximum gain within the configuration subspaces. 
The propagation scenario analyzed is far-field, with both direct and reflected paths present. We assume that the direct path is Rician distributed, while the \gls{RIS} reflected path is \gls{LOS}.
We show that the position can be used to determine a robust allocation strategy able to effectively utilize the designed codebook while keeping the \gls{OFDM} structure unchanged.
The performance is tested using max-rate and max-min allocation in both throughput and fairness and compared with the performance of a \gls{RIS}-aided system able to recover perfect \gls{CSI} through a minimum overhead \gls{CE} procedure.
Even considering the overhead obtained by the localization signaling needed to infer the position of the \gls{UE} in the system, we show that the proposed approach outperforms the \gls{CSI}-based system.\footnote{The simulation code for the paper is available at \url{https://github.com/lostinafro/ris-ofdm-loca-scheduling}}


\paragraph*{Notation} Integer sets are denoted by calligraphic letters $\mc{A}$ with cardinality $|\mc{A}|=A$. The complex Gaussian distribution is $\mc{CN}(\bm{\alpha},\mb{R})$ with mean $\bm{\alpha}$ and covariance matrix $\mb{R}$; non-central $\chi$-squared distribution with non-centrality $\xi$ and $v$ degrees of freedom is $\chi_v^{2}(\xi)$. Lowercase boldface letters denote column vectors $\mathbf{x}$; $\circ$ represents the element-wise vector multiplication. The identity matrix of size $N$ is $\mathbf{I}_N$, and $\mathbf{0}$ is a vector of zeros. 


\section{System model}
\label{sec:model}
We consider an \gls{UL} \gls{RIS}-aided communication scenario, where a single-antenna \gls{BS}, an \gls{RIS}, and set $\mc{K}$ of single-antenna \glspl{UE} are present. The \gls{BS} is assumed to have perfect knowledge of the \glspl{UE}' position, and it uses this information to make scheduling decisions.

\begin{figure}[t]
    \centering
    \includegraphics[width=6cm]{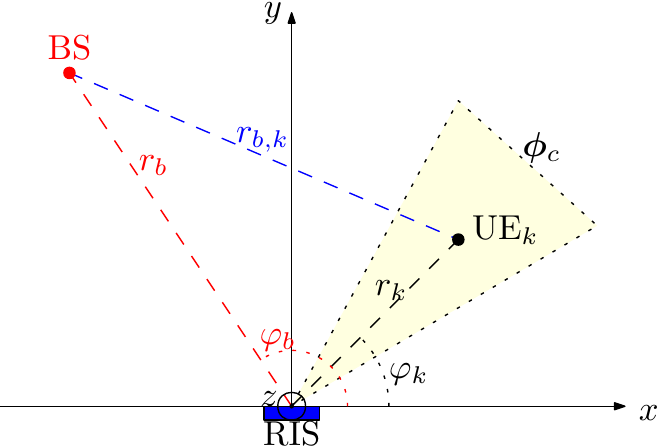}
    \caption{Scenario in 2D.}
    \label{fig:scenario}
\end{figure}

\paragraph{Geometry} 
The scenario is shown in Fig.~\ref{fig:scenario}, where the $z$ axis points to the outgoing direction of the $x-y$ plane. The center of the \gls{RIS} is positioned at the origin of the axis, parallel to $x$ and $z$ axis. The dimensions of the \gls{RIS} are $D_x$ and $D_z$, and it is formed by $N = N_x \times N_z$ elements, having distance $d_x = \frac{D_x}{N_x}$ and $d_z = \frac{D_z}{N_z}$, on the $x$ and $z$ axes respectively.  We enumerate each element following the direction of the $x$ and $z$ axis, i.e., the element $n,n'$ has center position $\mb{r}_{n,n'} = \left[ d_x\left(n - (N_x + 1) / 2\right), 0, d_z\left(n' - (N_z + 1)/2 \right)\right]\T$, $n = 1, \dots, N_x$, $n'=1,\dots, N_z$. 
The \gls{BS} is positioned at $\mb{r}_b = [r_b, \theta_b, \varphi_b]\T$. The users are collected in the set $\mc{K}$, $|\mc{K}| = K$. Each user $k\in\mc{K}$ is positioned at $\mb{r}_k = [r_k, \theta_k, \varphi_k]\T$. The distance between \gls{UE} $k$ and \gls{BS} is denoted as $r_{b,k} = || \mb{r}_b  - \mb{r}_k||$. 
For simplicity of presentation, the elevation angle of both \gls{BS} and \gls{UE} is set as $\theta_k = \theta_b = \pi/2$, i.e., \glspl{UE}, \gls{BS} and \gls{RIS} lay on the same plane. The design principles given here can be extended for different elevation angles.

\paragraph{RIS phase shift profile}
Each element $n,n'$ of the \gls{RIS} is able to imprint a 
phase shift $\phi_{n,n'}$ on the impinging electromagnetic wave\footnote{The \gls{RIS} is considered ideal, i.e., no attenuation or inter-element coupling is considered for mathematical tractability.}. The overall phase shift profile is a vector denoted as $\bm{\phi}\in\mathbb{C}^{N}$, collecting the phase shifts for each \gls{RIS}' element. A configuration \emph{codebook} collects the pre-determined phase shift profiles that can be loaded by the \gls{RIS}, i.e., \gls{RIS} and \gls{BS} share the knowledge of a finite set $\mc{C} = \{1, \dots, C\}$ of \gls{RIS}' configurations. When configuration $c\in\mc{C}$ is loaded, the correspondent \gls{RIS} phase shift profile $\bm{\phi}_c = [\phi_{1,1}^{(c)}, \dots, \phi_{N_x,1}^{(c)}, \phi_{1,2}^{(c)}, \dots, \phi_{N_x,2}^{(c)}, \dots, \phi_{N_x, N_z}^{(c)}]\T$ is designed to enhance the signal coming from a sector of the area, steering it towards the \gls{BS}. 

\paragraph{Proposed data-frame structure}
The time-frequency resource grid is formed by $S$ time slots, collected in the set $\mc{S} = \{0,1,\dots,S-1\}$, each of duration $T_s$, and $F$ \glspl{RB}, collected in the set $\mc{F}=\{0, \dots, F-1\}$, each having bandwidth $\Delta_F$.
The first \gls{RB} is assumed to have central frequency $f_0$, i.e., wavelength $\lambda_0 = \frac{\nu}{f_0}$, where $\nu$ denotes the propagation velocity of the wave. Hence, the wavelength of \gls{RB} $f \in\mc{F}$ is $\lambda_f = \frac{\nu}{f_0 + f \Delta_F}$.
It is assumed that the coherence time is longer than the overall frame duration in time ($S T_s$), while $\Delta_F$ is lower than the coherence bandwidth.

%
Each time slot comprises $\tau_\mathrm{OFDM}$ \gls{OFDM} symbols; $\tau_d \le \tau_\mathrm{OFDM}$ are reserved for data communication, while $\tau_\ell = \tau_\mathrm{OFDM} - \tau_d$ are reserved for localization waveform the \gls{BS} can use to keep track of the user position through a specific algorithm, such as~\cite{keykhosravi2021siso}. It is assumed that the \gls{BS} can recover the position of each user without any error\footnote{The design of the localization algorithm is out of the scope of the paper. Future works will focus on the trade-off of localization and communication performance in this setting, taking into account the impact of the localization error in the overall scheduling strategy.}. 

\begin{figure}
    \centering
    \includegraphics[width=0.9\columnwidth]{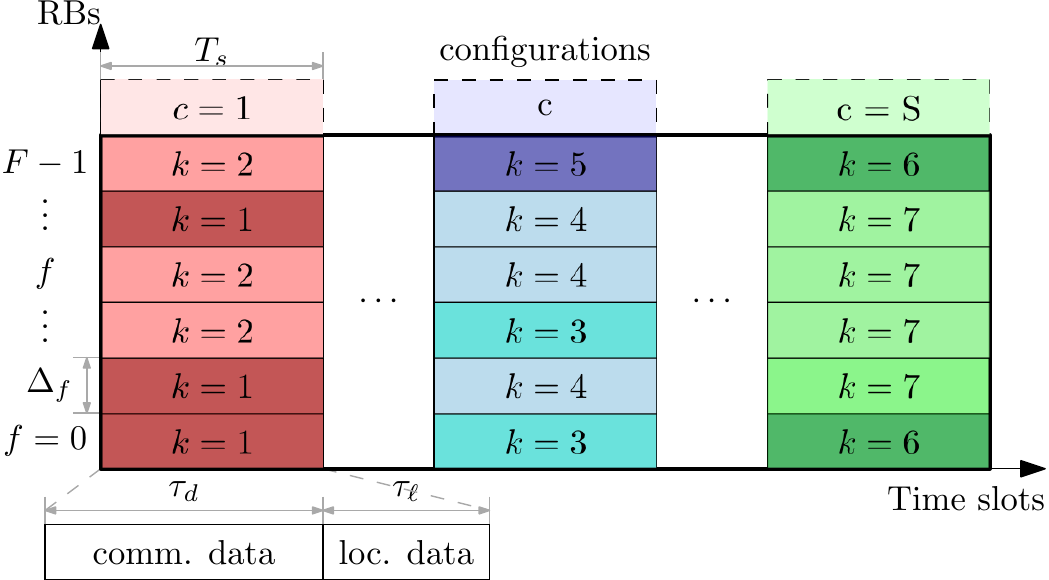}
    \caption{Example of configuration and time-frequency allocation for $K = 7$ users.} 
    \label{fig:dataframe}
\end{figure}

At the beginning of each time slot, the \gls{RIS} loads a single configuration $c$, which is kept for the whole time slot duration. 
It is assumed that the time for loading a configuration is lower than the time of the \gls{CP} of the first \gls{OFDM} symbol, and hence the \gls{RIS} switching through different configurations does not influence the transmitted data (differently from, e.g.,~\cite{croisfelt2022oracle}).
\glspl{UE} are multiplexed through orthogonal slot-\gls{RB} allocation, as shown in the time-frequency example given in Fig.~\ref{fig:dataframe}. 

\paragraph{Signal model}
\label{sec:signal}
Consider that a single time-frequency resource related to configuration $c\in\mc{C}$ and frequency $f\in\mc{F}$ is exclusively allocated to user $k\in\mc{K}$. On that resource, the user transmits a symbols $x \in \mathbb{C}$, $\E{x} = 0$, $\E{|x|^2} = 1$, with power $P_k$. The received signal at the \gls{BS} is given by
\begin{equation}
    y_{k,f,c} = \Big( h_{k,f} + g_{k,f}(\bm{\phi}_c) \Big) \sqrt{P_k} x + w
\end{equation}
where we indicated $h_{f,k}\in \mathbb{C}$ as the direct \gls{BS}-\gls{UE} channel, $g_{f,k}(\bm{\phi}_c)\in\mathbb{C}$ as the \gls{BS}-\gls{RIS}-\gls{UE} reflected channel while the \gls{RIS} has configuration $c$ loaded, and $w \sim \mc{CN}(0, \sigma^2)$ as the \gls{AWGN} at the \gls{BS} receiver side.
The \gls{SNR} at the \gls{RB} is given by
\begin{equation} \label{eq:snr}
    \gamma_{k,f,c} = \frac{P_k}{\sigma^2} \left| |h_{k,f}|e^{j\angle h_{k,f}} + |g_{k,f}(\bm{\phi}_c)|e^{j\angle g_{k,f}(\bm{\phi}_c)} \right|^2
\end{equation}
where $\angle h_{k,f}$ is the phase shift that the signal experiences through the direct path, while $\angle g_{k,f}(\bm{\phi}_c)$ is the phase shift experienced through the reflective path comprehending the one imprinted by the \gls{RIS} configuration $c$. The \gls{SNR} of a single resource is maximized when $\angle g_{k,f}(\mb{\phi}_c) = \angle  h_{k,f}$. In the following, the assumption made on the channels and their formulation is given.

\begin{assumption} \label{assu:channel}
In this work, the direct channel coefficient is assumed to be Rician distributed, i.e., it has a deterministic \gls{LOS} component and a random \gls{NLOS} component; the reflected path has only a dominant deterministic \gls{LOS} component. \end{assumption}

According to Assumption~\ref{assu:channel}, the vector $\mb{h}_k, \mb{g}_k \in\mathbb{C}^F$ collecting the channel of the direct and reflective paths for all the \glspl{RB} are
\begin{align}
    &\mb{h}_{k} = \sqrt{\beta_k(r_{b,k})} \left[\sqrt{\frac{\kappa}{\kappa + 1}} \mb{d}(r_{b,k})   + \sqrt{\frac{1}{\kappa + 1}} \mb{n}_{k} \right], \label{eq:channel:direct}\\
    &\mb{g}_k(\bm{\phi}_c) = \hspace{-0.5mm}\sqrt{\beta_k(r_k r_b)} N \mb{d}(r_b + r_k) \hspace{-.8mm}\circ\hspace{-.5mm}\mb{a}_k(\bm{\phi}_c),  \label{eq:channel:reflected}
\end{align}
where $\mb{n}_k \sim \mc{CN}(\mb{0}, \mb{I}_F)$ is the random \gls{NLOS} component; the deterministic \gls{LOS} component is represented by
\begin{equation}
    \mb{d}(r) = e^{-j \frac{2 \pi}{\nu} f_0 r}  [1, e^{-j \frac{2 \pi}{\nu} \Delta_F r}, \dots, e^{-j \frac{2 \pi}{\nu} (F-1) \Delta_F r}]\T
\end{equation}
which collects the different propagation delays for different \glspl{RB}; $\kappa$ is the Rician factor previously estimated in the environment~\cite{Greenstein1999kfactor}; the path loss is~\cite{albanese2022marisa}
\begin{equation} \label{eq:pathloss}
    \beta_k(r) = \beta_0 \frac{G_k G_b}{r^\varepsilon} 
\end{equation}
having path-loss exponent $\varepsilon$\footnote{Here, we used the common assumption that the variation of the path loss on the different wavelengths is negligible.} and \gls{BS} and \gls{UE} antenna gains $G_b$ and $G_k$, respectively; finally, $\mb{a}_k(\bm{\phi}_c)$ is the normalized \gls{AF} generated by configuration $c$. 

Remark that \glspl{UE}, \gls{RIS} and \gls{BS} lay on the same plane; to provide the maximum gain on the $x-y$ plane, the phase shifts induced by the \gls{RIS} elements in the $z$-dimension are set to the same~\cite{Balanis2012antenna}. In other words, $\phi_{n,n'}^{(c)} = \phi_n$, $n'=1,\dots, N_z$, $\forall c \in\mc{C}$. The contribution of the normalized \gls{AF} on a frequency $f\in\mc{F}$ results~\cite{Balanis2012antenna, tang2020wireless}
\begin{equation} \label{eq:af:RB}
\begin{aligned} 
    a_{k, f}(\bm{\phi}_c) =& \, \frac{N_z}{N} e^{-j \frac{2 \pi}{\nu} (f_0 + f \Delta_F) \frac{N_x+1}{2} d_x (\cos\varphi_k + \cos\varphi_b)} \\
    \quad& \sum_{n=1}^{N_x} e^{j\phi_{n}^{(c)}} e^{j \frac{2 \pi}{\nu} (f_0 + f \Delta_F) n d_x (\cos\varphi_k + \cos\varphi_b)}.
\end{aligned}
\end{equation}

\section{Resource grid and codebook design}
\label{sec:codebook}
In this section, we propose a design of the codebook to cover the whole area of interest.
We remark that the element phase shift term $\phi_n^{(c)}$ influences all the \glspl{RB} in the same manner\footnote{Circuital-based \gls{RIS} models show that the \gls{RIS} phase shift is affected by the frequency~\cite{Li2021widebandris}. We neglect this effect to concisely show the design principle of the proposed protocol, considering that this dependency is deterministic and can be straightforwardly included.}; hence, we will design the phase shift for a single \gls{RB}, e.g., $f=0$, study the impact on the other frequencies, and finally provide a design of the configuration codebook to employ for the proposed paradigm.

\subsection{General phase-shift formulation}
For each configuration $c\in\mc{C}$, we aim to maximize the \gls{AF} gain when the received signal comes from a specific angular direction $\varphi_c$, for $f = 0$.  
By the observation of eq.~\eqref{eq:af:RB}, we can infer that each element phase shift $\phi_n^{(c)}$ needs to compensate for the phase shift generated by the \gls{BS} position, while let each term of the sum to be 1 when $\varphi_k = \varphi_c$.
Accordingly, we can design the phase shifts as
\begin{equation} \label{eq:phi_n}
    \phi_{n}^{(c)} = \frac{2 \pi}{\nu} f_0 (\phi_\text{res}^{(c)} - n d_x \phi^{(c)}_x),
\end{equation}
with
\begin{equation} \label{eq:phaseshifts}
\begin{cases}
   \phi_\text{res}^{(c)} = (\cos\varphi_b + \cos\varphi_c) \frac{N_x + 1}{2} d_x, \\
    \phi_{x}^{(c)} = \cos\varphi_b + \cos\varphi_c,
\end{cases}
\end{equation}
being $\phi_\text{res}^{(c)}$ set to cancel the residual phase shift given by the geometry of the \gls{RIS}. Indeed, plugging~\eqref{eq:phi_n}-\eqref{eq:phaseshifts} into eq.~\eqref{eq:af:RB} yields
\begin{equation} \label{eq:af:linearphase}
\begin{aligned}
    a_{k,f}(\bm{\phi}_c) &=  e^{-j \frac{2 \pi}{\nu} \frac{N_x+1}{2} d_x \left[f_0  (\cos\varphi_k - \cos\varphi_c) + f \Delta_F (\cos\varphi_k + \cos\varphi_b) \right]} \\
    & \frac{1}{N_x} \sum_{n=1}^{N_x} e^{j \frac{2 \pi}{\nu} n  d_x \left[f_0 (\cos\varphi_k - \cos\varphi_c) + f \Delta_F (\cos\varphi_k+ \cos\varphi_b) \right]}, \\
\end{aligned}
\end{equation}
which is maximized for $f=0$ if a user is located at $\varphi_k = \varphi_c$.
For $f \neq 0$, this design can remove the contribution of the \gls{AF} in the overall phase of the reflective channel 
$\angle g_{k,f}(\bm{\phi}_c)$. Indeed, by means of finite geometric series,  eq.~\eqref{eq:af:linearphase} can be rewritten as~\cite{Balanis2012antenna, tang2020wireless}
\begin{equation} \label{eq:af:sin}
\begin{aligned}
    a_{k,f}(\bm{\phi}_c) =& \frac{\sin\left(\frac{\pi d_x}{\nu} N_x \psi_{k,f} \right)}{N_x \sin\left(\frac{\pi d_x}{\nu} \psi_{k,f} \right)}
\end{aligned}
\end{equation}
where $\psi_{k,f} = f_0 (\cos\varphi_k - \cos\varphi_c) + f \Delta_F (\cos\varphi_k + \cos\varphi_b)$.
Eq.~\eqref{eq:af:sin} shows that $a_{k,f}(\bm{\phi}_c) \in \mathbb{R}$, i.e., $\angle a_{k,f}(\bm{\phi}_c) = \pi u(-a_{k,f}(\bm{\phi}_c))$ where $u(\cdot)$ is the Heaviside step function. Moreover, eq.~\eqref{eq:af:sin} is the \gls{AF} of the uniform linear phased array, being negative only outside of the main lobe region~\cite{Balanis2012antenna}. 
Hence, as long as the allocated configuration and \glspl{RB} guarantee that the normalized \gls{AF} works in the main lobe for the user $k$, the contribution of the \gls{AF} in the total phase $\angle g_{k,f}(\bm{\phi}_k)$ is zero, i.e., $\angle a_{k,f}(\bm{\phi}_c) = 0$.

\begin{figure}[th]
    \centering
    \input{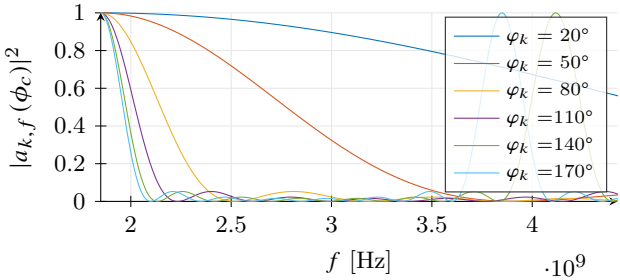}
    \caption{\gls{AF} vs. the frequency and for various values of $\varphi_k$, where $\varphi_c = \varphi_k$. $N_x = 8$, $d_x = d_z = \lambda_0 / 2$ m, and $\varphi_b = 150$°.}
    \label{fig:af:freq:phik}
\end{figure}

\subsection{Frequency dependency}
Regarding the influence on the frequency domain, we can prove experimentally that the \gls{AF} varies slowly with the different \glspl{RB}. A depiction of the frequency dependence of the \gls{AF} is given in Fig.~\ref{fig:af:freq:phik}; the plot is made for different angular positions and assuming that $\phi_c = \phi_k$, $\forall k$, to show the frequency dependence only. It can be seen that the position of the nulls of the function change according to the user position: the higher the azimuth angle, the narrower the main lobe\footnote{Remark that for particular positions $\cos\phi_k = -\cos\phi_b$, the \gls{AF} never changes, regardless the employed \gls{RB}.}.
Nevertheless, even considering the user at $\varphi_k = 170$°, to obtain a reduction of 20\% of the array gain, a frequency hop of $63.36$ MHz from the first \gls{RB} must be realized, i.e, $352$ \glspl{RB} with numerology 0~\cite{3gpp:rel15:MAC}. Therefore, we assume that the system has a number of available \glspl{RB} $F \le 350$ to minimize the effect of the variation of the \gls{AF} gain while guaranteeing that the contribution on the phase provided by the design~\eqref{eq:phi_n} is zero, regardless of the selected \gls{RB}.

\subsection{Codebook design}
\begin{figure}[thb]
    \centering
    \input{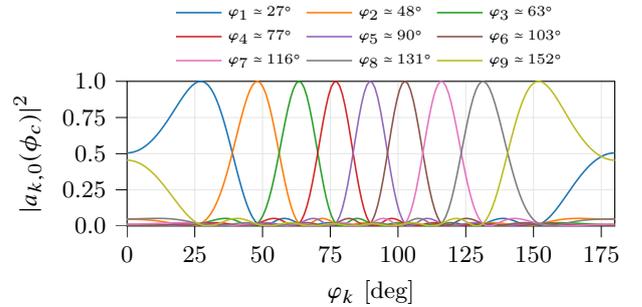}
    \caption{Visualization of the \gls{AF} using the codebook design in~\eqref{eq:codebook-design}, with $\tau = 0.5$, $x_\tau = 1.391$, and $N_x=8$.}
    \label{fig:codebook-design}
\end{figure}

To cover the whole area of interest using the proposed framework, we resort to an approach that enables us to create a codebook containing at least one configuration providing an \gls{AF} gain of at least $\tau \in (0,1]$ to any \gls{UE} $k$ regardless of its position, for $f=0$. To do so, we design a codebook having every consequent configuration to overlap at the angular direction that provides the desired minimum gain, similar to~\cite{croisfelt2022oracle}. 
For the sake of simplicity, we set $\tau = -3$ dB, but any other values can be considered with the same procedure. With negligible error, we approximate the main lobe of eq.~\eqref{eq:af:sin} by a $\sinc$ function~\cite{Balanis2012antenna, croisfelt2022oracle}. Thus, we find the argument such that $|\sinc(x_\tau)|^2 = \tau = 0.5$, which gives the value of $x_\tau \approx \pm 1.391$. Therefore, $\sinc(\pi d_x / \nu N_x \psi_{k,0}) \ge \tau$ when
\[
- x_\tau \le \frac{\pi d_x}{\nu} N_x f_0 (\cos\varphi_k - \cos\varphi_c) \le x_\tau,
\]
Now, define as $\varphi\HPm_c$ and $\varphi\HPp_c$ the angular directions where the \gls{AF} gain for configuration $c$ is half of the maximum, where $\varphi\HPm_c < \varphi_c < \varphi\HPp_c$; to let consequent configurations having overlapping half power angular direction, we impose $\varphi\HPm_{c+1} = \varphi\HPp_{c}$, $\forall c \in\mc{C}$. Accordingly, the set of equation to satisfy is $\forall c \in\mc{C}$
\begin{equation} \label{eq:halfpowersystem}
    \begin{cases}
        \frac{\pi d_x}{\nu} N_x f_0 \left( \cos\varphi\HPm_c - \cos\varphi_c \right) = x_\tau, \\
        \frac{\pi d_x}{\nu} N_x f_0 \left( \cos\varphi\HPp_c - \cos\varphi_c \right) = -x_\tau, \\
        \varphi\HPm_{c+1} = \varphi\HPp_{c}.
    \end{cases}
\end{equation}
To cover the whole area, we impose that $\varphi\HPm_1 = 0$, meaning that the first configuration has the half power direction towards the right limit of the area of interest. Therefore, by solving~\eqref{eq:halfpowersystem} in an iterative manner, we obtain
\begin{equation} \label{eq:codebook-design}
\begin{cases}
    \cos\varphi_c = 1 - (2 c - 1)  \frac{\nu x_\tau}{\pi d_x N_x f_0}, \\
    \cos\varphi_c\HPm = 1 - (2 c - 2)  \frac{\nu x_\tau}{\pi d_x N_x f_0}, \\
    \cos\varphi_c\HPp = 1 - (2 c)  \frac{\nu x_\tau}{\pi d_x N_x f_0},
\end{cases}
\end{equation}
where the number of configurations to cover the area is
\begin{equation}
        C = \min\left\{c\in\mathbb{N} \,|\, \cos\varphi\HPp_c < -1\right\} = \left\lceil \frac{\pi d_x N_x f_0}{x_\tau \nu} \right\rceil.
    \label{eq:codebook-design:C}
\end{equation}
A visualization of the obtained codebook is given in Fig.~\ref{fig:codebook-design}.

\section{Resource allocation} \label{sec:scheduling}
In this section, we present the allocation scheduling process based on the proposed framework.

Given the set of configurations $\mc{C}$, the set of available of \glspl{RB}, and the position information of each user, 
we know the value of the deterministic part of the \gls{SNR} for all the \glspl{UE}, \glspl{RB} and time-slots. Nevertheless, the actual \gls{SNR} is a random variable: to provide a feasible communication rate, the \gls{SE} set for the data transmission must be lower than the achievable capacity. 
To overcome this issue, we define the $\epsilon$-robust \gls{SE} $r_{k,f,c}^\epsilon$ as the \gls{SE} that assures that the data transmitted by \gls{UE} $k$ on \gls{RB} $f$ and under configuration $c\in\mc{C}$ are successfully decoded with probability $\epsilon$, i.e., the \gls{SE} satisfying
\begin{equation}
    \mc{P}\left[r_{k,f,c}^\epsilon \le \log_2\left( 1 + \gamma_{k,f,c}\right) \right] \le 1 - \epsilon.
\end{equation}
\begin{lemma}
Following Assumption~\ref{assu:channel}, the  $\epsilon$-robust \gls{SE} is
\begin{equation}
    r_{k,f,c}^\epsilon = \log_2\left(1 + \frac{P_k}{2\sigma^2(\kappa +1)} F^{-1}_{\tilde{\gamma}_{k,f,c}}(1 - \epsilon) \right)
\end{equation}
where $F^{-1}_{\tilde{\gamma}_{k,f,c}}(\cdot)$ is the inverse \gls{CDF} function of $\tilde{\gamma}_{k,f,c} \sim \chi_2^{2}(\xi_{k,f,c})$, where $\xi_{k,f,c}$ is given in eq.~\eqref{eq:nc} at the top of next page.
\end{lemma}
\begin{figure*}[bt]
    \centering
\begin{equation} \label{eq:nc}
\xi_{k,f,c} = 2\left(\kappa + \frac{\kappa + 1}{\beta(r_{b,k})} |g_{f,k}(\mb{\phi}_c)|^2 + 2 \sqrt{\frac{\kappa + 1}{\beta(r_{b,k})}} \Re\{ d_{f}(r_{k,b}) g_{k,f}(\mb{\phi}_c)\} \right)
\end{equation}
\par\noindent\rule{\textwidth}{0.4pt}
\end{figure*}
\begin{proof}
The proof comes directly from the definition of the channels~\eqref{eq:channel:direct}-\eqref{eq:channel:reflected} after few manipulations. 
\end{proof}

In practice, knowing the deterministic part of the \gls{SNR} allows us to compute the $\epsilon$-robust \gls{SE} for all the resources and use it to optimize the allocation.
To this purpose, we further define the allocation variable $\rho_{k,f,c}\in\{0,1\}$, mapping each user with the corresponding time-frequency resource. Accordingly, the total \gls{SE} of user $k$ is
\begin{equation} \label{eq:rk}
    r_k = \sum_{c\in\mc{C}} \sum_{f\in\mc{F}} \rho_{k,f,c} r_{k,f,c}^\epsilon
\end{equation}
To evaluate the system performance with different metrics, we focus on two classical problems: maximize the overall rate, and maximize the minimum rate.

\paragraph{Max-Rate}
In order to maximize the overall communication throughput, we define the following optimization problem:
\begin{align}
    \max_{\bm{\rho}} & \sum_{k\in\mc{K}} r_k \label{eq:max-rate}\\    
    \text{s.t.} \quad & \rho_{k,f,c} \in \{0,1\}, \tag{\ref{eq:max-rate}.a} \quad \forall f \in \mc{F}, \label{eq:max-rate:integer}\\
    & \sum_{k\in\mc{K}} \rho_{k,f,c} \le 1, \tag{\ref{eq:max-rate}.b} \quad \forall f \in \mc{F}, \, \forall c\in\mc{C}, \label{eq:max-rate:rho}
\end{align}
where constraint~\eqref{eq:max-rate:integer} enforces the integer condition on the allocation variable, and constraint~\eqref{eq:max-rate:rho} assures that each time-frequency resource is allocated to at most one user.

The optimal solution of problem~\eqref{eq:max-rate} is trivial: each \gls{RB} of each slot is given to the \gls{UE} providing the highest $\epsilon$-robust \gls{SE}, i.e.,
\begin{equation} \label{eq:allovar}
    \rho_{k,f,c} =
    \begin{cases}
    1, \quad f = \argmax_{k\in\mc{K}} r_{k,f,c}^\epsilon,  \\
    0, \quad\text{othewise},
    \end{cases}
    \, \forall c\in\mc{C}.
\end{equation}

\paragraph{Max-min}
In order to maximize the minimum rate between the users, we define the following optimization problem:
\begin{equation} \label{eq:max-min}
\begin{aligned}
    &\max_{\bm{\rho}} \min_{k\in\mc{K}}~r_k  
    &\text{s.t} \quad \eqref{eq:max-rate:integer}, \, \eqref{eq:max-rate:rho}.
\end{aligned}    
\end{equation}

In this case, the problem is not trivial, and an optimal solution is hard to achieve. Therefore, we resort to a heuristic scheme that iteratively allocates the \gls{RB} to the \gls{UE} with the smallest rate~\cite{Radunovic2007maxmin}. The procedure is given in Algorithm~\ref{alg:max-min}, where $A = F C$, and $a$ represents the double index $(f,c)$ in lexicographic order.

\setlength{\textfloatsep}{0pt}
\begin{algorithm}[t]
\caption{Max-min allocation}
\label{alg:max-min}
\footnotesize
 \textbf{Initialize}: $r_k = 0$, $\forall k \in\mc{K}$, $\mc{A} = \{1, \dots, A\}$\; 
 \tcc{Assign at least one \gls{RB}}
   \For{$k\in\mc{K}$}{
    $\hat{a} = \argmax_{a\in\mc{A}} r_{k,:,:}^\epsilon$\;
    $\rho_{k,\hat{a}} = 1$\;     
    $\mc{A} = \mc{A} \setminus \{\hat{a}\}$\;
    update $r_k$ by~\eqref{eq:rk}\;
   }
   \tcc{Distribute the remaining \glspl{RB}}
   \While{$\mc{A} \neq \empty$}{
   $\hat{k} = \argmin_{k\in\mc{K}} r_k$\;
    $\hat{a} = \argmax_{a\in\mc{A}} r_{\hat{k},:,:}^\epsilon$\;
    $\rho_{\hat{k},\hat{a}} = 1$\;
    update $r_k$ by~\eqref{eq:rk}\;
   }   
\end{algorithm}

\subsection{Joint vs. sequential allocation}
The scheduling procedure described performs both the $\epsilon$-robust rates evaluation and the allocation operation on the entire dimension of the tensor $K \times F \times C$. While the optimization problems presented here are relatively simple, performing more complicated resource allocation schemes may be computationally infeasible in a real application. 
In those cases, we can resort to a sequential allocation: time slot allocation first, followed by \gls{RB} allocation.

The time slot allocation is determined by comparing the angular position of the user with the angular pointing of the various configuration.
Denoting as $c_k$ the configuration employed by user $k$, we have
\begin{equation} \label{eq:conf-allocation}
    c_k = \argmin_c \left\{ |\varphi_c - \varphi_k|\right\},
\end{equation}
and the time slot allocation is obtained straightforwardly. 

For the \gls{RB} allocation, let us denote as $\mc{K}_c \subseteq \mc{K}$ the subset of the users allocated to configuration $c\in\mc{C}$. Consequently, we can apply eq.~\eqref{eq:allovar} for a max-rate allocation, taking care of substituting $\mc{K}_c$ to $\mc{K}$ in the rule.
Similarly, Algorithm~\ref{alg:max-min} can be applied for max-min allocations, where now $A = F$, and $a$ represents the index~$f$.

\section{Numerical results} \label{sec:results}

\setlength{\textfloatsep}{15pt}
\begin{table}[tb]
    \centering
    \caption{Simulation parameters.}\vspace{-0.2cm}
    \footnotesize
    {\renewcommand{\arraystretch}{1}
    \begin{tabular}{@{}lcc@{}}
    \toprule
    Parameter & Symbol & Value \\ \midrule
    Outer/inner radius & $R_\mathrm{out}, R_\mathrm{inn}$ & $30/9$~m\\     
    BS position (Cartesian) & $\mb{r}_b$ & $[10, 100, 0]\T$~m\\
    RIS element spacing & $d_x = d_z $ & $\lambda / 2$\\
    Number of RIS elements & $N_x = N_z$ & 10\\    
     Central frequency & $f_0$ & 1.8 GHz\\
     Bandwidth & $\Delta_F$ & $180$~kHz\\
     \glspl{RB} & $F$ & 50 \\
     Time slots/Configurations & $S=C$ & $11$\\
     OFDM symbols & $\tau_\mathrm{OFDM}$ & 14 \\
     OFDM data/localization symbols & $\tau_d, \tau_\ell$ & 7, 7 \\
     Success probability & $\epsilon$ & $95$\% \\     
     Antenna gains &$G_b \cdot G_k$ & $12.85$~dB\\     
     Path loss exponent &$\varepsilon$ &$2.7$\\
     Reference path gain & $\beta_0$ & $-31.53$~dB\\     
     \bottomrule
    \end{tabular}\vspace{-0.4cm}}
    \label{tab:params}
\end{table}

In this section, we present our numerical results. 
To perform the simulations, we have deployed the $K$ users randomly in a ring having inner radius $R_\mathrm{inn}$ and outer radius $R_\mathrm{out}$. The parameters we used are listed in Table~\ref{tab:params}.
As a benchmark, we consider the \gls{CSI}-based approach working as follows: before the communication data phase, minimal length pilot sequences~\cite{Wang2020ce} are sent by the \glspl{UE} to perform \gls{CE}, one complex symbol per \gls{OFDM} symbol to cover all the \glspl{RB}. The estimation is assumed to be \emph{noise-free}, i.e., the \gls{BS} has perfect \gls{CSI} after the \gls{CE} ends. In the following, we label \texttt{csi} the performance obtained by the \gls{CSI}-based scheme, while \texttt{jnt} and \texttt{seq} the localization-based joint and sequential allocation performance.
The main performance metrics used are the average total throughput obtained by the system, and Jain's fairness index, given by
\begin{equation} \label{eq:metrics}
\begin{aligned}
    \bar{r} = \eta^{(i)} \frac{\Delta_F}{S} \sum_{k\in\mc{K}} r_k, \quad
    J = \frac{\left(\sum_{k\in\mc{K}} r_k\right)^2}{\sum_{k\in\mc{K}} r_k^2}
\end{aligned}
\end{equation}
In eq.~\eqref{eq:metrics}, $\eta^{(i)}$ represent the efficiency of the $i$-th communication scheme given by $\eta^{(\texttt{jnt})} = \eta^{(\texttt{seq})} = \epsilon \frac{\tau_d}{\tau_d + \tau_\ell}$ and $\eta^{(\texttt{csi})} =  \frac{S \tau_\mathrm{OFDM}}{S \tau_\mathrm{OFDM} + \tau_\mathrm{csi}}$,
where $\tau_\mathrm{csi} = \lceil K (N_x + 1)\rceil$~\cite{Wang2020ce}.

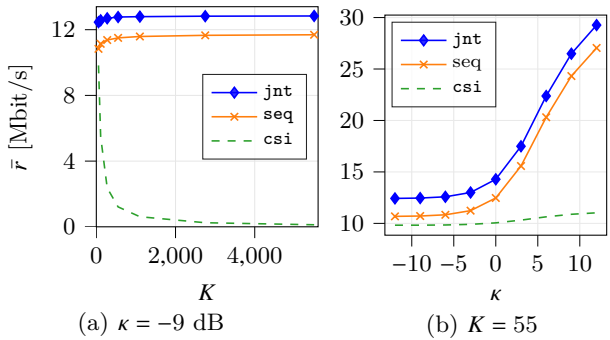
\begin{figure}[thb]
    \centering
    \begin{subfigure}[t]{.45\columnwidth}
    \centering
    \input{img/tikz/max_rate_throughput_vs_K.tex}
    \vspace{-0.6cm}
    \caption{$\kappa = -9$ dB}
    \end{subfigure}
    ~
    \begin{subfigure}[t]{.45\columnwidth}
    \input{img/tikz/max_rate_throughput_vs_kappa.tex}
    \vspace{-0.15cm}
    \caption{$K = 55$}
    \end{subfigure}
    \caption{Max-rate performance.}
    \label{fig:max-rate}
\end{figure}

Fig.~\ref{fig:max-rate} shows the performance obtained by the max-rate allocation schemes. Fig.~\ref{fig:max-rate}(a) presents the throughput as a function of the number of users $K$ when $\kappa = -9$ dB, while Fig.~\ref{fig:max-rate}(b) as a function of the Rician $\kappa$ factor when $K = 55$. Both localization-based paradigms outperform the CSI-based scheme. In particular, when the system is in an overloaded state, i.e., $K > F S$, the \gls{CE} overhead increases in a prohibitive manner. Interestingly, even with a low number of users, the localization-based schemes achieve superior performance for every value of $\kappa$ under test.

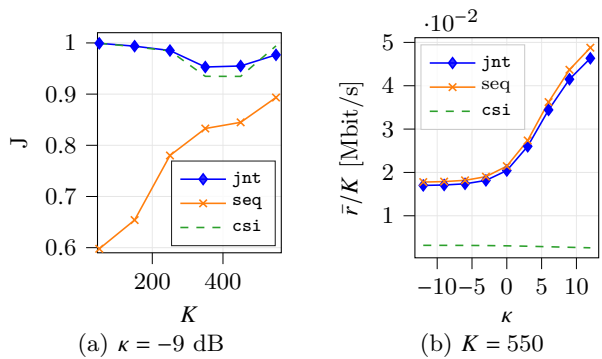
\begin{figure}[thb]
    \centering
    \begin{subfigure}[t]{.45\columnwidth}
    \centering
    \input{img/tikz/max_min_fairness_vs_K_kappa-9.tex}
    \vspace{-0.2cm}
    \caption{$\kappa = -9$ dB}
    \end{subfigure}
    ~
    \begin{subfigure}[t]{.45\columnwidth}
    \input{img/tikz/max_min_throughput_vs_kappa.tex}
    \vspace{-0.2cm}
    \caption{$K = 550$}
    \end{subfigure}
    \caption{Min-max performance.}
    \label{fig:min-max}
\end{figure}

Fig.~\ref{fig:min-max} shows the performance of the min-max allocation schemes. Fig.~\ref{fig:min-max}(a) presents the fairness as a function of the number of users $K$ when $\kappa = -9$ dB, while Fig.~\ref{fig:min-max}(b) the average throughput per user as a function of the Rician $\kappa$ factor when $K = 550$. In this case, the fairness performance of \texttt{jnt} and the \texttt{csi} are comparable, both outperforming \texttt{seq} for every $K$. Regarding the average throughput per user, the localization-based schemes outperform \texttt{csi}. In particular, \texttt{jnt} has a clear advantage to \texttt{csi} reaching similar fairness, but with an average rate more than doubled.

\section{Conclusions} \label{sec:conclusions}
This paper proposed an \gls{OFDM} framework for \gls{RIS}-aided communications that exploit localization information to perform a robust resource allocation in a multi-user scenario. The results presented show the effectiveness of the proposed approach w.r.t. \gls{CE}-based solutions.
While the proposed solution is promising, the impact of localization accuracy needs to be studied by investigating the trade-off of transmitting data and localization symbols.

\bibliographystyle{IEEEtran}
\bibliography{loca.bib}
\end{document}

%% file: acronym.tex
\newacronym{AWGN}{AWGN}{additive white gaussian noise}
\newacronym{AF}{AF}{array factor}
\newacronym{BS}{BS}{base station}

\newacronym{CE}{CE}{channel estimation}
\newacronym{CDF}{CDF}{cumulative distribution function}
\newacronym{CP}{CP}{cyclic prefix}
\newacronym{CMS}{CMS}{canonical minimum-scattering}
\newacronym{CSI}{CSI}{channel state information}

\newacronym{DFT}{DFT}{discrete fourier transform}
\newacronym{DTFT}{DTFT}{discrete time fourier transform}
\newacronym{DMA}{DMA}{dynamic metasurface antenna}
\newacronym{DL}{DL}{downlink}

\newacronym{ELAA}{ELAA}{extremely large aperture array}

\newacronym{flop}{flop}{floating point operation}
\newacronym{flops}{flops}{\gls{flop} per second}
\newacronym{FD}{FD}{full-digital}
\newacronym{FDMA}{FDMA}{Frequency division multiple access}

\newacronym{LOS}{LoS}{Line-of-Sight}

\newacronym{NLOS}{NLoS}{Non-Line-of-Sight}

\newacronym{ISAC}{ISAC}{integrated sensing and communication}

\newacronym{OMP}{OMP}{orthogonal matching pursuit}
\newacronym{OFDM}{OFDM}{orthogonal-frequency division multiplexing}

\newacronym{MINLP}{MINLP}{mixed integer non-linear problem}

\newacronym{RF}{RF}{radio-frequency}
\newacronym{RIS}{RIS}{reconfigurable intelligent surface}
\newacronym{RB}{RB}{resource block}

\newacronym{SE}{SE}{spectral efficiency}
\newacronym{SINR}{SINR}{signal-to-interference-plus-noise ratio}
\newacronym{SNR}{SNR}{signal-to-noise ratio}
\newacronym{SLL}{SLL}{side lobe level}

\newacronym{TDD}{TDD}{time division duplexing}
\newacronym{TDMA}{TDMA}{time division multiple access}

\newacronym{UE}{UE}{user equipment}
\newacronym{UL}{UL}{uplink}
\newacronym{ULA}{ULA}{uniform linear array}

\newacronym{VR}{VR}{visibility region}



%% file: img/tikz/max_rate_throughput_vs_K.tex
\begin{tikzpicture}

\definecolor{darkorange25512714}{RGB}{255,127,14}
\definecolor{forestgreen4416044}{RGB}{44,160,44}
\definecolor{lavender233}{RGB}{233,233,233}
\definecolor{lightgray204}{RGB}{204,204,204}
\definecolor{steelblue31119180}{RGB}{31,119,180}

\begin{axis}[
height=4.5cm,
width=4.5cm,
legend cell align={left},
legend style={
  at={(0.95,0.5)},
  anchor=east,
},
tick align=outside,
tick pos=left,
xlabel={\(\displaystyle K\)},
xmajorgrids,
xmin=0, xmax=5600,
ylabel={$\bar{r}$ [Mbit/s]},
ymajorgrids,
ymin=0, ymax=13.4,
ytick={0,4,8,12,16},
yticklabels={
  \(\displaystyle {0}\),  
  \(\displaystyle {4}\),  
  \(\displaystyle {8}\),  
  \(\displaystyle {12}\),  
  \(\displaystyle {16}\)
}
]
\addplot [semithick, blue, mark=diamond*, mark size=2, mark options={solid}]
table {%
55 12.4591638807284
110 12.5701536588773
275 12.6935160461787
550 12.7707188464804
1100 12.7940235664421
2750 12.8223251573294
5500 12.8347220016408
};
\addlegendentry{\texttt{jnt}}
\addplot [semithick, darkorange25512714, mark=x, mark size=2, mark options={solid}]
table {%
55 10.833729732105
110 11.1402765877564
275 11.3707952474629
550 11.5018852764772
1100 11.5867252518525
2750 11.6571920560123
5500 11.6912288290793
};
\addlegendentry{\texttt{seq}}
\addplot [semithick, forestgreen4416044, dashed]
table {%
55 9.82513393032722
110 5.47815589153312
275 2.36519895283018
550 1.21233467996489
1100 0.613706141716637
2750 0.248056895571638
5500 0.124500295417333
};
\addlegendentry{\texttt{csi}}


\end{axis}
\end{tikzpicture}

%% file: img/tikz/max_rate_throughput_vs_kappa.tex
\begin{tikzpicture}

\definecolor{darkorange25512714}{RGB}{255,127,14}
\definecolor{forestgreen4416044}{RGB}{44,160,44}
\definecolor{lavender233}{RGB}{233,233,233}
\definecolor{lightgray204}{RGB}{204,204,204}
\definecolor{steelblue31119180}{RGB}{31,119,180}

\begin{axis}[
width=4.5cm,
height=4.5cm,
legend cell align={left},
legend style={
  at={(0.03,0.97)},
  anchor=north west,
},
tick align=outside,
tick pos=left,
xlabel={\(\displaystyle \kappa\)},
xmajorgrids,
xmin=-13.2, xmax=13.2,
xtick style={color=black},
xtick={-15,-10,-5,0,5,10,15},
xticklabels={
  \(\displaystyle {\ensuremath{-}15}\),
  \(\displaystyle {\ensuremath{-}10}\),
  \(\displaystyle {\ensuremath{-}5}\),
  \(\displaystyle {0}\),
  \(\displaystyle {5}\),
  \(\displaystyle {10}\),
  \(\displaystyle {15}\)
},
y grid style={lavender233},
ymajorgrids,
ymin=8.85151856515297, ymax=30.2401732554369,
]
\addplot [semithick, blue, mark=diamond*, mark size=2, mark options={solid}]
table {%
-12 12.4253106471137
-9 12.4591638807284
-6 12.582139971413
-3 13.0008775582797
0 14.2841671286686
3 17.4916631555555
6 22.3769478059178
9 26.486869304421
12 29.2679616786058
};
\addlegendentry{\texttt{jnt}}
\addplot [semithick, darkorange25512714, mark=x, mark size=2, mark options={solid}]
table {%
-12 10.6815308282906
-9 10.7148854519774
-6 10.833729732105
-3 11.2353867488805
0 12.4677671770795
3 15.5742720350395
6 20.3118302410626
9 24.3174212523339
12 27.0461267749496
};
\addlegendentry{seq}
\addplot [semithick, forestgreen4416044, dashed]
table {%
-12 9.82373014198406
-9 9.82513393032722
-6 9.84433921682666
-3 9.90570822272558
0 10.0498429168162
3 10.3158027189047
6 10.6369015559041
9 10.8840136701281
12 11.0271133097424
};
\addlegendentry{\texttt{csi}}
\end{axis}

\end{tikzpicture}

%% file: img/tikz/max_min_fairness_vs_K_kappa-9.tex
\begin{tikzpicture}

\definecolor{darkorange25512714}{RGB}{255,127,14}
\definecolor{forestgreen4416044}{RGB}{44,160,44}
\definecolor{lavender233}{RGB}{233,233,233}
\definecolor{lightgray204}{RGB}{204,204,204}
\definecolor{steelblue31119180}{RGB}{31,119,180}

\begin{axis}[
width=4cm,
height=4.5cm,
legend cell align={left},
legend style={    
  at={(0.97,0.03)},
  anchor=south east,  
},
tick align=outside,
tick pos=left,
x grid style={lavender233},
xlabel={\(\displaystyle K\)},
xmajorgrids,
xmin=45, xmax=565,
ylabel={J},
ymajorgrids,
ymin=0.59, ymax=1.02,
]
\addplot [semithick, blue, mark=diamond*, mark size=2, mark options={solid}]
table {%
50 0.999243175182723
150 0.993687367218431
250 0.985143131611637
350 0.952952148422242
450 0.954962947982503
550 0.976677116758078
};
\addlegendentry{\texttt{jnt}}
\addplot [semithick, darkorange25512714, mark=x, mark size=2, mark options={solid}]
table {%
50 0.597780849098348
150 0.653956906662436
250 0.780085171473543
350 0.832804255726962
450 0.844830327950017
550 0.893363834964714
};
\addlegendentry{\texttt{seq}}
\addplot [semithick, forestgreen4416044, dashed]
table {%
50 0.99954520334938
150 0.992487928448371
250 0.984377203977808
350 0.934768614966926
450 0.934606977131258
550 0.994244305910316
};
\addlegendentry{\texttt{csi}}
\end{axis}

\end{tikzpicture}

%% file: img/tikz/max_min_throughput_vs_kappa.tex
\begin{tikzpicture}

\definecolor{darkorange25512714}{RGB}{255,127,14}
\definecolor{forestgreen4416044}{RGB}{44,160,44}
\definecolor{lavender233}{RGB}{233,233,233}
\definecolor{lightgray204}{RGB}{204,204,204}
\definecolor{steelblue31119180}{RGB}{31,119,180}

\begin{axis}[
width=4cm,
height=4.5cm,
legend cell align={left},
legend style={
  fill opacity=0.8,
  draw opacity=1,
  text opacity=1,
  at={(0.03,0.97)},
  anchor=north west,
  draw=lightgray204
},
tick align=outside,
tick pos=left,
x grid style={lavender233},
xlabel={\(\displaystyle \kappa\)},
xmajorgrids,
xmin=-13.2, xmax=13.2,
xtick style={color=black},
xtick={-15,-10,-5,0,5,10,15},
xticklabels={
  \(\displaystyle {\ensuremath{-}15}\),
  \(\displaystyle {\ensuremath{-}10}\),
  \(\displaystyle {\ensuremath{-}5}\),
  \(\displaystyle {0}\),
  \(\displaystyle {5}\),
  \(\displaystyle {10}\),
  \(\displaystyle {15}\)
},
ylabel={$\bar{r} / K$ [Mbit/s]},
ymajorgrids,
ymin=0.000300957579279767, ymax=0.0510836953745609,
ytick={0,0.01,0.02,0.03,0.04,0.05,0.06},
yticklabels={
  \(\displaystyle {0}\),
  \(\displaystyle {1}\),
  \(\displaystyle {2}\),
  \(\displaystyle {3}\),
  \(\displaystyle {4}\),
  \(\displaystyle {5}\),
  \(\displaystyle {6}\)
}
]
\addplot [semithick, blue, mark=diamond*, mark size=2, mark options={solid}]
table {%
-12 0.017000251648027
-9 0.0171010964796189
-6 0.0173696568632962
-3 0.0181577285664528
0 0.0204351952958227
3 0.0260241969197762
6 0.0344228229243866
9 0.0414986771419747
12 0.0463304628823666
};
\addlegendentry{\texttt{jnt}}
\addplot [semithick, darkorange25512714, mark=x, mark size=2, mark options={solid}]
table {%
-12 0.0177732295279551
-9 0.0178857928732386
-6 0.018178492661637
-3 0.0190242933067968
0 0.0214514239847763
3 0.0273758130754146
6 0.0362343673946174
9 0.0436866553580576
12 0.048775389111139
};
\addlegendentry{seq}
\addplot [semithick, forestgreen4416044, dashed]
table {%
-12 0.00317995938227601
-9 0.00317558316780354
-6 0.00316296850567068
-3 0.00313080517468929
0 0.00306783547733286
3 0.00297246980402384
6 0.00285335995232013
9 0.00272741471281966
12 0.00260926384270163
};
\addlegendentry{\texttt{csi}}
\end{axis}

\end{tikzpicture}

%% file: conference_paper.bbl
\begin{thebibliography}{10}
\providecommand{\url}[1]{#1}
\csname url@samestyle\endcsname
\providecommand{\newblock}{\relax}
\providecommand{\bibinfo}[2]{#2}
\providecommand{\BIBentrySTDinterwordspacing}{\spaceskip=0pt\relax}
\providecommand{\BIBentryALTinterwordstretchfactor}{4}
\providecommand{\BIBentryALTinterwordspacing}{\spaceskip=\fontdimen2\font plus
\BIBentryALTinterwordstretchfactor\fontdimen3\font minus
  \fontdimen4\font\relax}
\providecommand{\BIBforeignlanguage}[2]{{%
\expandafter\ifx\csname l@#1\endcsname\relax
\typeout{** WARNING: IEEEtran.bst: No hyphenation pattern has been}%
\typeout{** loaded for the language `#1'. Using the pattern for}%
\typeout{** the default language instead.}%
\else
\language=\csname l@#1\endcsname
\fi
#2}}
\providecommand{\BIBdecl}{\relax}
\BIBdecl

\bibitem{bjornson2021signalprocessing}
E.~Björnson,  \emph{et~al.}, ``Reconfigurable intelligent surfaces: A signal
  processing perspective with wireless applications,'' \emph{IEEE Signal
  Processing Mag.}, vol.~39, no.~2, pp. 135--158, 2022.

\bibitem{mengnan2022survey}
M.~Jian \emph{et~al.}, ``Reconfigurable intelligent surfaces for wireless
  communications: Overview of hardware designs, channel models, and estimation
  techniques,'' \emph{Intelligent and Converged Networks}, vol.~3, no.~1, pp.
  1--32, 2022.

\bibitem{Zheng2020cebf}
B.~Zheng and R.~Zhang, ``Intelligent reflecting surface-enhanced ofdm: Channel
  estimation and reflection optimization,'' \emph{IEEE Wireless Communications
  Letters}, vol.~9, no.~4, pp. 518--522, 2020.

\bibitem{Wang2020ce}
Z.~Wang \emph{et~al.}, ``Channel estimation for intelligent reflecting surface
  assisted multiuser communications: Framework, algorithms, and analysis,''
  \emph{IEEE Transactions on Wireless Communications}, vol.~19, no.~10, pp.
  6607--6620, 2020.

\bibitem{keykhosravi2021siso}
K.~Keykhosravi, M.~F. Keskin, G.~Seco-Granados, and H.~Wymeersch, ``{SISO}
  {RIS}-enabled joint 3d downlink localization and synchronization,'' in
  \emph{ICC 2021-IEEE International Conference on Communications}.\hskip 1em
  plus 0.5em minus 0.4em\relax IEEE, 2021, pp. 1--6.

\bibitem{Abrardo2021positioning}
A.~Abrardo \emph{et~al.}, ``Intelligent reflecting surfaces: Sum-rate
  optimization based on statistical position information,'' \emph{IEEE Trans.
  on Communications}, vol.~69, no.~10, pp. 7121--7136, 2021.

\bibitem{wang2021joint}
R.~Wang, Z.~Xing, and E.~Liu, ``Joint location and communication study for
  intelligent reflecting surface aided wireless communication system,''
  \emph{arXiv preprint arXiv:2103.01063}, 2021.

\bibitem{croisfelt2022oracle}
\BIBentryALTinterwordspacing
V.~Croisfelt \emph{et~al.}, ``Random access protocol with channel oracle
  enabled by a reconfigurable intelligent surface,'' 2022. [Online]. Available:
  \url{https://arxiv.org/abs/2210.04230}
\BIBentrySTDinterwordspacing

\bibitem{Greenstein1999kfactor}
L.~Greenstein \emph{et~al.}, ``Moment-method estimation of the ricean
  k-factor,'' \emph{IEEE Communications Letters}, vol.~3, no.~6, pp. 175--176,
  1999.

\bibitem{albanese2022marisa}
A.~Albanese, F.~Devoti, V.~Sciancalepore \emph{et~al.}, ``{MARISA: A
  self-configuring metasurfaces absorption and reflection solution towards
  6G},'' in \emph{IEEE INFOCOM 2022-IEEE Conference on Computer
  Communications}.\hskip 1em plus 0.5em minus 0.4em\relax IEEE, 2022, pp.
  250--259.

\bibitem{Balanis2012antenna}
C.~A. Balanis, \emph{Antenna theory: analysis and design}.\hskip 1em plus 0.5em
  minus 0.4em\relax Wiley-Interscience, 2005.

\bibitem{tang2020wireless}
W.~Tang \emph{et~al.}, ``Wireless communications with reconfigurable
  intelligent surface: Path loss modeling and experimental measurement,''
  \emph{IEEE Transactions on Wireless Communications}, vol.~20, no.~1, pp.
  421--439, 2020.

\bibitem{Li2021widebandris}
H.~Li \emph{et~al.}, ``Intelligent reflecting surface enhanced wideband
  {MIMO-OFDM} communications: From practical model to reflection
  optimization,'' \emph{IEEE Transactions on Communications}, vol.~69, no.~7,
  pp. 4807--4820, 2021.

\bibitem{3gpp:rel15:MAC}
3GPP, ``{Medium Access Control (MAC) protocol specification},'' {3rd Generation
  Partnership Project (3GPP)}, Technical Report (TR) 38.321, 10 2018, version
  15.2.0.

\bibitem{Radunovic2007maxmin}
B.~Radunovic \emph{et~al.}, ``A unified framework for max-min and min-max
  fairness with applications,'' \emph{IEEE/ACM Transactions on Networking},
  vol.~15, no.~5, pp. 1073--1083, 2007.

\end{thebibliography}
